\documentclass[aps,pra,superscriptaddress,reprint,floatfix,notitlepage,nobalancelastpage]{revtex4-1}
\usepackage{graphicx,amsmath,amssymb,bm,color,engord,nicefrac,bbold,mathtools,braket}
\usepackage[usenames,dvipsnames]{xcolor}
\usepackage[caption=false]{subfig}
\usepackage[colorlinks,colorlinks,urlcolor=Magenta,citecolor=Magenta,linkcolor=black]{hyperref}
\DeclareMathAlphabet\mathbfcal{OMS}{cmsy}{b}{n}

\begin{document}

\title{Hartree-Fock-Bogoliubov theory of trapped one-dimensional imbalanced Fermi systems}
\author{Kelly R. Patton}
\email[\hspace{-1.4mm}]{kpatton@georgiasouthern.edu}
\affiliation{Department of Physics \& Astronomy, Georgia Southern University-Armstrong Campus, Savannah, GA 31419, USA}
\author{Daniel E. Sheehy}
\affiliation{Department of Physics \& Astronomy, Louisiana State University, Baton Rouge, Louisiana 70803, USA}
\date{\today}
\begin{abstract} Ground state Hartree-Fock-Bogoliubov (HFB) theory is
applied to  imbalanced spin-$\nicefrac{1}{2}$  one-dimensional Fermi
systems that are spatially confined by either a harmonic or a
hard-wall trapping potential.
It has been hoped that such systems,
which can be realized using ultracold atomic gases,  would exhibit the
long-sought-after  Fulde-Ferrell-Larkin-Ovchinnikov (FFLO) superfluid
phase. The HFB formalism   generalizes the standard Bogoliubov
quasi-particle transformation, by allowing for Cooper  pairing to exist
between all possible single-particle states, and accounts for
the effects of the inhomogeneous trapping potential as well as  the
mean-field Hartree potential.
This provides an unbiased framework to describe inhomgenous densities and pairing correlations in the
FFLO state of a confined 1D gas.    In  a harmonic trap, numerical minimization
of the HFB ground state energy yields a spatially oscillating order parameter reminiscent of the FFLO state.  However,
we find that this state has almost no imprint in the local fermion densities (consistent with experiments that found
no evidence of the FFLO phase).  
In contrast, for a hard-wall
geometry, we find a strong signature of the spatial oscillations of
the FFLO pairing amplitude reflected in the local {\it in situ}
densities. In the hard wall case, the excess spins are strongly localized near regions
where there is a node in the pairing amplitude, creating an
unmistakeable crystalline modulation of the density.
\end{abstract}
\maketitle

\section{Introduction}
Irrespective of the temperature, applying a sufficiently strong
external magnetic field to a  superconductor destroys the
superconducting state. For a conventional superconductor,
 a spin-imbalanced Fermi liquid
becomes the energetically favored  phase
above this critical  field strength.   Driving this phase
transition is the mismatch of the spin-$\uparrow$ and
spin-$\downarrow$  Fermi energies, which is a consequence  of the
Zeeman splitting caused by the applied magnetic field. Within
Bardeen Cooper Schrieffer (BCS) theory  \cite{BCS}, for an $s$-wave
superconductor and weak interactions, once the Zeeman  splitting
reaches a critical value, on the order of the 
superconducting  gap $\Delta$, the
so-called Chandrasekhar-Clogston limit
\cite{ClogstonPRL62,ChandrasekharAPL62}, the superconducting state is
no longer energetically favorable. Depending on the effective
dimensionality and interaction strength, a similar phase transition
can occur in neutral fermionic superfluids, which are now commonly
realized using ultracold atomic gases. In such systems, the 
mismatch of Fermi energies is accomplished by selectively populating
two pseudo-spin-$\nicefrac{1}{2}$ hyperfine states of the atoms. Thus, in
contrast with an archetypal condensed matter system, the spin
imbalance in ultracold atomic gases can be readily fine-tuned across
the full parameter range, from zero imbalance, equal numbers of each
spin, to fully polarized.

\begin{figure}[ht]\hspace{.3cm}
\parbox[h]{8.0cm}{
\includegraphics*[width=\linewidth,angle=0]{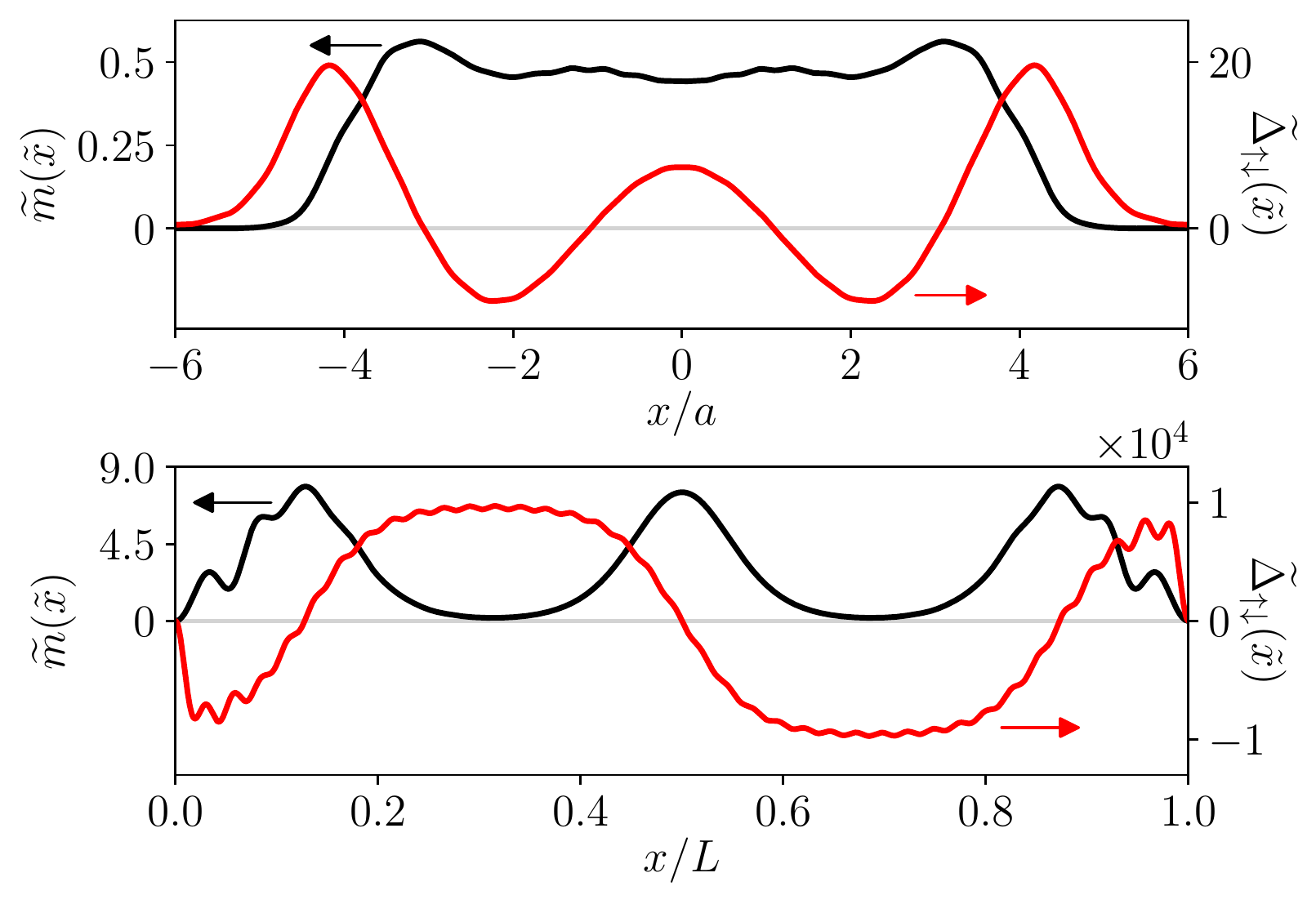}}\hspace{.5cm}
\parbox[h]{8cm}{\caption{(Color online) The top and bottom panels each show
    the local magnetization (black curves)  and pairing amplitude (red curves) for
    a one-dimensional imbalanced Fermi superfluid confined in a parabolic single-particle potential (top panel)
    and a homogeneous box trap (bottom panel).  While each case shows oscillatory pairing correlations reminiscent of
    an FFLO~\cite{FuldePR64,LarkinJETP64} phase, the signature of such pairing in the local density is much
    stronger in the case of a box trap. The arrows indicate which $y$-axis each curve corresponds to.  (All parameters are the same as in Figs.~\ref{fig:fig1} and \ref{fig:fig2}.)
}
\label{fig:nought}}
\end{figure} 

The Fermi liquid phase is not the only possibility for fields greater than the Chandrasekhar-Clogston
limit.  Indeed, 
several  unconventional  superconducting/fluid  phases have been
theoretically predicted to exist in this regime.
One such  phase is the
Fulde-Ferrell-Larkin-Ovchinnikov (FFLO) state \cite{FuldePR64,LarkinJETP64}, which has attracted considerable attention for many years. Unlike a
conventional fermionic superfluid, where the Cooper pairs have zero net
 momentum, the FFLO state is comprised of  pairs having 
 momentum ${\bm Q}$.  In a translationally  invariant  system
and at  weak coupling, it can be shown that ${\bm Q}\approx {\bm k}^{\uparrow}_{\rm F}- {\bm
  k}^{\downarrow}_{\rm F}$, where ${\bm k}^{\sigma}_{\rm F}$ are the
Fermi wave vectors of each spin.  This  momentum corresponds  to a  periodic real space order parameter $\Delta({\bm r})=\Delta({\bm
  r}+{\bm a})$, where the lattice constant $|{\bm a}|\approx
2\pi/|{\bm Q}|$. Fulde and Ferrell (FF) proposed a plane wave order
parameter  $\Delta({\bm r})\propto {\rm e}^{i{\bm {\bm Q}\cdot {\bm r}}}$, while
Larkin and Ovchinnikov (LO) proposed a standing wave version  $\Delta({\bm r})\propto
\cos({\bm {\bm Q}\cdot {\bm r})}$.  Above the
Chandrasekhar-Clogston limit,  both are  energetically more favorable
than a  Fermi liquid, but the LO state is believed to have the lowest energy of the two.

In addition to the ground state spontaneously  breaking the $U(1)$
symmetry related to fixed particle number, the FF-state order
parameter would also break inversion symmetry, while the LO state would
break translational invariance. It is thought that the breaking of
these additional symmetries, especially for an LO state, would lead to
a clear  experimental signature in the local density
\cite{NovelSuperfluidsBook}.   Indeed,  an LO-type
superfluid  would possess a modulated density that is commensurate
with the oscillations of the local pairing function.  Physically, this
density modulation results from the  unpaired atoms, due to the
imbalance,  localizing near the  nodes of the  pairing
amplitude. As a result,  the simultaneous coexistence of
a magnetic lattice order and a superfluid  would occur.   Unfortunately, despite  considerable effort over the past sixty years,
little to no conclusive experimental evidence of an FFLO state in
either ultracold gases \cite{Zwierlein2006,Partridge2006,Shin2006,
Partridge2006prl, Navon2010,Olsen2015,Revelle2016,
Sommer2012,Zhang2012,Ries,Murthy,Boettcher,Fenech,Cheng2016,Mitra2016,
LiaoNature2010} or  condensed matter systems
\cite{MayaffreNatPhys2014, PrestigiacomoPRB2014} has been found. 

As previously mentioned, the particle number of each pseudo-spin is externally  controllable in
ultracold atomic systems, as well as the
atoms' effective spatial dimensionality and  inter-particle
interactions. These experimental controls  make  ultracold atomic
gases  an almost ideal physical system to realize exotic states of
matter.
While the regime of stability for the FFLO state of a trapped Fermi gas
is predicted to be rather
narrow in three dimensions~\cite{Sheehy06,Parish07,SR2007}, the situation
improves in lower dimensions.  
In particular,  the FFLO phase
has been theoretically  predicted to be stable over a wide parameter range in 
 one dimension (1D)
\cite{OrsoPRL07,HuPRL07}.   Experiments  \cite{LiaoNature2010} for a
harmonically trapped gas in 1D  show that at small population
imbalances,  the gas is  locally magnetized only in a central region
of the trap, while the edges of the cloud remain unpolarized. The
spatial extent of the polarized central region grows with increasing
imbalance until a critical polarization  $P_{\rm c}$ is reached, above
which the entire cloud becomes magnetic. Nevertheless, no sign of an
underlying FFLO-like order parameter is discernible in the {\it in
situ} densities. However, it is unclear how the anisotropic trapping
potential, which is omnipresent in ultracold atomic systems and breaks
translational symmetry,  effects spatially varying phases, such as
FFLO, or their detection.  

Further uncertainty arises as predictions from different  theoretical
methods  have varying degrees of agreement with each other and the
experiments.  For example, exact thermodynamic Bethe ansatz combined with the
 local density approximation (BA+LDA) appears to be
consistent with experimental results \cite{OrsoPRL07,HuPRL07,
  Kakashvili2009,ZhaoPRL2009, GuanRMP2013}. The results of BA+LDA
predict a magnetized central core that grows with increasing
polarization. The critical polarization $P_{\rm c}$,  above which the
entire cloud is magnetized, is also in general agreement with the
experimental results. In principle, the Bethe ansatz gives the exact many-body wave function, from which the local density can be obtained. Unfortunately, extracting the local density
from the many-body wave function is difficult.   Thus, the results of
Refs. \onlinecite{OrsoPRL07,HuPRL07, Kakashvili2009,ZhaoPRL2009,
  GuanRMP2013} only represent  the {\it average} density (total particle number $N$ per volume $V$: $n=N/V$) in the thermodynamic
limit and not the exact  local density $n(x)$.  The average  density  in the
inhomogeneous trap is found within the LDA, which amounts to replacing the thermodynamic chemical potential  ($\mu$) dependence of the average density, $n(\mu)$, by a spatially vary one, $\mu\to \mu-V(x)$, where $V(x)$ is the trapping potential; $n( \mu-V(x))$.  Unsurprisingly, the results
show no spatial modulation of the average density that would be
indicative of an FFLO-like state.  To directly take into account the
effects of the trapping potential, the present authors put forth a
BCS-like variational wave function in
Ref.~\onlinecite{PattonPRA2017}. Reminiscent of the experiment and
BA+LDA, this wave function also produced  magnetized and unmagnetized
regions, but, unlike BA+LDA, it further showed LO-like oscillations in
the local pairing amplitude. Nonetheless, no signature of these
oscillations was reflected in the local {\it in situ} densities.   In
contrast, local mean-field theory
\cite{LiuPRA2007,LiuPRA2008,LuPRL,SunPRL2012}  and various lattice
models and methods
\cite{FeiguinPRB2007,PoliniPRB2008,TezukaPRL2008,BatrouniPRL2008,MolinaPRL2009,
  BakhtiariPRL2008} tend to show  large oscillations in the pairing
amplitude  that are clearly (in most cases) correlated with a
modulation of the local density or magnetization.

In this article, we take yet another approach to the physics of FFLO phases in trapped fermionic atomic
gases with an imposed population imbalance. We apply the
configuration based Hartree-Fock-Bogoliubov (HFB) \cite{SchuckBook,
  AdvancesInNuclearPhysicsBook,BassemNuPhysA,DukelskyJPG85} theory to
these systems. HFB theory is a generalization of Hartree-Fock
mean-field theory to systems with  Cooper pairing. Unlike standard BCS theory, where the Bogoliubov
quasi-particles are a  linear combination of a single particle and a
single hole, in HFB theory, a quasi-particle  is represented by a
linear combination  of all possible particle and hole states. This is
of importance because, in a balanced and translationally invariant
system, the only particle-particle interaction terms that are meaningful  in the
renormalization-group (RG) sense are the ones that give rise to the formation of
standard  Cooper pairs, i.e., the interaction between plane-wave-time-reserved
states \cite{PolchinshiArxiv}. However, since the universality class of an infinite FFLO system is currently unknown,   it is not clear what  the
relevant interactions  (in the RG sense) in a trapped
and/or imbalanced system are. Hartree-Fock-Bogoliubov theory circumvents this ambiguity by allowing for a broad range of pairing and density correlations. 

 An additional advantage of the HFB approach is that it
 takes
into account the  effect of the nontrivial inhomogeneous Hartree
potential.  The real space, or coordinate, version of HFB has
previously  been applied to trapped and imbalanced systems
\cite{LiuPRA2007,LiuPRA2008,LuPRL,SunPRL2012}. Here, we apply HFB in
the single-particle basis, the so-called configuration formalism.  This formalism  has been highly  developed in the nuclear
physics community \cite{SchuckBook}.  The benefit  of this approach is that it results in
more  detailed  information about the system. For example,
besides the local densities and pairing amplitude, one has access to
the occupation probability of each level, the pairing amplitude
between all single-particle states, the mode resolved single-particle
density matrix, or any other equal-time ground state correlation function.  This leads to building a richer and more physically intuitive
picture of the FFLO state in trapped systems.   Additionally, this
method allows for optimal control over the size of the  Hilbert space
needed for numerical calculations as the strength of particle-particle interactions
is increased.  This will become especially important in
higher-dimensional systems where regularization of the two-body interaction potential is necessary.  Applying HFB to ultracold atomic gases in higher dimensions will be part of future work.

In the following sections, we present the application of HFB theory to find the mean-field ground state of polarized
spin-\nicefrac{1}{2}  fermions in 1D.  We obtained  results for two specific systems:  
Fermions with a harmonic (parabolic) confining trap   and  fermions with a box-shaped (hard-wall, or ``homogeneous'')
confining potential
\cite{ZwierleinPRL2017,HueckPRL2018}.
In the harmonic case, our HFB method agrees with both experiments and BA+LDA theory for observables like the critical
$P_{\rm c}$, giving us confidence in this method.  Our main findings concern the existence and nature (and
potential observability) of any FFLO pairing correlations in the presence of these two types of trapping potential.  
In general, we find that the
ground states of both systems, in the polarized regime, are  FFLO-like, showing spatial
oscillations in both the local pairing amplitude and the local
densities. However, the details show significant differences between the two
systems as shown in Fig.~\ref{fig:nought}.  For the harmonically trapped system, the amplitudes of the
density modulations are relatively small, especially in comparison to
the hard-wall case, and would probably be entirely washed out at finite
temperature \cite{LiuPRA2008} or by other experimental limitations (such as imaging resolution).
Another striking difference is that unlike the harmonic trap, there is
no central region of magnetization in the hard-wall system. Instead,
at all polarizations, the system is only magnetic near the nodes of
the pairing amplitude with  zero magnetization elsewhere, i.e., the
local magnetization has a definitive crystalline order.  Thus, our results imply
that it may be much easier to 
experimentally detect the FFLO state in a box-shaped trap.

The rest of this Paper is organized as follows.  In Sec.~\ref{Sec: Theory}, we present the general HFB theory for a system of fermions
in one spatial dimension subject to a trapping potential $V(x)$.  In Sec.~\ref{Sec: Results} we apply the HFB theory to the two specific
systems described above, i.e., the case of harmonically trapped atoms characterized by trap frequency $\omega_0$
(i.e., $V(x) = \frac{1}{2}m\omega_0^2 x^2$) and the case of a homogeneous box
trap of size $L$ (i.e., $V(x) = 0$ for $0<x<L$ and $V(x) = \infty$ elsewhere).
In 
Sec.~\ref{Sec: Discussion and Conclusions}, we elaborate on these
results and  discuss the  prospects of their extension and future
work. 
\section{Theory}
\label{Sec: Theory}
Here, for completeness, we recap the salient  aspects of HFB theory,
which can  be  found in the literature \cite{SchuckBook,
  AdvancesInNuclearPhysicsBook,BassemNuPhysA,DukelskyJPG85}, although,
unfortunately,  with varying notational conventions.    In principle,
this  can be applied to systems of arbitrary dimensionality, but
currently we will only  be  interested in 1D systems.  

We take the second quantized 1D Hamiltonian of a trapped interacting spin-\nicefrac{1}{2} Fermi system to be $(\hbar = 1)$:
\begin{align}
\label{Exact Hamiltonian}
\hat{H}&= \sum_{\sigma}\int\limits_{-\infty}^{\infty} {\rm d}x\, \hat{\Psi}^{\dagger}_{\sigma}(x)\left[-\frac{\nabla^{2}}{2m}+V(x)\right]\hat{\Psi}^{}_{\sigma}(x)\\ \nonumber&\mkern-18mu+\frac{1}{2}\sum_{\sigma,\sigma'}\int\limits_{-\infty}^{\infty}{\rm d}x{\rm d}x'\,  \hat{\Psi}^{\dagger}_{\sigma}(x) \hat{\Psi}^{\dagger}_{\sigma'}(x')U(x-x') \hat{\Psi}^{}_{\sigma'}(x')\hat{\Psi}^{}_{\sigma}(x),
\end{align}
where $V(x)$ is the external trapping potential, and $U(x-x')$ is the two-body interaction.
In the following sections we will restrict ourselves to  a harmonic or hard-wall trapping potential and assume an attractive short-range interaction,
which is common for ultracold atomic gases, but for now the trapping potential and interaction will remain arbitrary. 

Expanding the field operators in terms of mode operators, $ \hat{\Psi}^{(\dagger)}_{\sigma}(x)=\sum_{n}\psi^{(*)}_{n}(x)\hat{a}^{(\dagger)}_{n\sigma}$, the Hamiltonian becomes
\begin{equation}
\label{Exact Hamiltonian: spin-quantum number index}
\hat{H}=\sum_{n,\sigma}\epsilon^{}_{n}\hat{a}^{\dagger}_{n\sigma}\hat{a}^{}_{n\sigma}+\frac{1}{2}\sum_{\sigma,\sigma'}\sum_{i,j,k,l}U^{}_{i,j,k,l}\hat{a}^{\dagger}_{i\sigma}\hat{a}^{\dagger}_{j\sigma'}\hat{a}^{}_{k\sigma'}\hat{a}^{}_{l\sigma},
\end{equation}
where
\begin{equation}
\label{single-particle spectrum}
\int\limits_{-\infty}^{\infty} {\rm d}x\, \psi^{*}_{n}(x)\left[-\frac{\nabla^{2}}{2m}+V(x)\right]\psi_{n'}(x)=\delta^{}_{n,n'}\epsilon^{}_{n},
\end{equation}
and
\begin{equation}
\label{pseudo two-body interaction matrix element}
U^{}_{i,j,k,l}=\int\limits_{-\infty}^{\infty}{\rm d}x{\rm d}x'\, \psi^{*}_{i}(x)\psi^{*}_{j}(x')U(x-x')\psi^{}_{k}(x')\psi^{}_{l}(x).
\end{equation}
Going forward, it will be convenient to express the terms appearing in Eq.~\eqref{Exact Hamiltonian: spin-quantum number index}  using  a composite orbital-spin index; $\alpha = (n,\sigma)$. Then, the matrix elements of the kinetic energy and interaction terms  are
\begin{equation}
T^{}_{\alpha,\beta} = \delta^{}_{n^{}_{\alpha},n^{}_{\beta}}\delta^{}_{\sigma^{}_{\alpha},\sigma^{}_{\beta}}\epsilon^{}_{n_{\alpha}}=\delta^{}_{\alpha, \beta}\epsilon^{}_{\alpha},
\end{equation}
and  
\begin{equation}
U^{}_{\alpha,\beta,\delta,\gamma}=\delta^{}_{\sigma^{}_{\alpha},\sigma^{}_{\gamma}}\delta^{}_{\sigma^{}_{\beta},\sigma^{}_{\delta}}U^{}_{i_{\alpha},j_{\beta},k_{\delta},l_{\gamma}}.
\end{equation}
The full Hamiltonian is, then,
\begin{equation}
\label{Exact Hamiltonian: composite index}
\hat{H}=\sum_{\alpha,\beta}T^{}_{\alpha,\beta}\hat{a}^{\dagger}_{\alpha}\hat{a}^{}_{\beta}+\frac{1}{2}\sum_{\alpha,\beta,\delta,\gamma}U^{}_{\alpha,\beta,\delta,\gamma}\hat{a}^{\dagger}_{\alpha}\hat{a}^{\dagger}_{\beta}\hat{a}^{}_{\delta}\hat{a}^{}_{\gamma},
\end{equation}
where $\sum_{\alpha}=\sum_{n}\sum_{\sigma}$ etc.
It will be further convenient to anti-symmetrize the interaction term,  $U^{}_{\alpha,\beta,\delta,\gamma}$, in the last two indices: 
\begin{equation}
U^{}_{\alpha,\beta,[\delta,\gamma]}=\frac{1}{2}\left(U^{}_{\alpha,\beta,\delta,\gamma}-U^{}_{\alpha,\beta,\gamma,\delta}\right)\equiv \frac{1}{2}\overline{U}^{}_{\alpha,\beta,\delta,\gamma}.
\end{equation}
 Using Eq.~\eqref{pseudo two-body interaction matrix element} one can  show $\overline{U}^{}_{\alpha,\beta,\delta,\gamma}$ has the following symmetry relations:
\begin{equation}
\label{symmetry relations of Uoverline}
\overline{U}^{}_{\alpha,\beta,\delta,\gamma}=-\overline{U}^{}_{\alpha,\beta,\gamma,\delta}=-\overline{U}^{}_{\beta,\alpha,\delta,\gamma} = \overline{U}^{}_{\beta,\alpha,\gamma,\delta} = \overline{U}^{*}_{\delta,\gamma,\alpha,\beta}
\end{equation}
This finally gives~\cite{footnote}

\begin{equation}
\label{Full Hamiltonian in composite index notation}
\hat{H}=\sum_{\alpha,\beta}T^{}_{\alpha,\beta}\hat{a}^{\dagger}_{\alpha}\hat{a}^{}_{\beta}+\frac{1}{4}\sum_{\alpha,\beta,\delta,\gamma}\overline{U}^{}_{\alpha,\beta,\delta,\gamma}\hat{a}^{\dagger}_{\alpha}\hat{a}^{\dagger}_{\beta}\hat{a}^{}_{\delta}\hat{a}^{}_{\gamma}.
\end{equation}

The basic idea of HFB theory is to generalize the well-known Bogoliubov-Valatin transformation \cite{BogoliubovNuovoCimento,ValatinNuovoCimento}    to allow for the most general transformation to  quasi-particle operators, $\hat{\gamma}^{}_{\alpha}$ and $\hat{\gamma}^{\dagger}_{\alpha}$,  in terms of the single-particle states.  To do this we
define:
\begin{subequations}
\label{quasi-particle  transformation}
\begin{align}
\hat{\gamma}^{\dagger}_{\alpha}=\sum_{\alpha'}\left[\left({\bm V}^{\rm T}\right)^{}_{\alpha,\alpha'}\hat{a}^{}_{\alpha'}+\left(\bm{U}^{\rm T}\right)^{}_{\alpha,\alpha'}\hat{a}^{\dagger}_{\alpha'}\right],\\
\hat{\gamma}^{}_{\alpha}=\sum_{\alpha'}\left[\left({\bm V}^{\dagger}\right)^{}_{\alpha,\alpha'}\hat{a}^{\dagger}_{\alpha'}+\left(\bm{U}^{\dagger}\right)_{\alpha,\alpha'}\hat{a}^{}_{\alpha'}\right],
\end{align}
\end{subequations}
where ${\bm V}$ and ${\bm U}$ are matrices whose elements are the variational parameters that will  used to minimize the mean-field ground state energy.  The inverse transformation from  quasi-particle back to single-particle operators can be most easily found by first  defining the column vector
\begin{equation}
\left(\begin{array}{c}{\bm{\hat{a}^{\phantom{\dagger}}}} \\{\bm{\hat{a}}^{\dagger}}\end{array}\right)=\left(\begin{array}{c}\hat{a}_{n_{1}\uparrow} \\\hat{a}_{n_{2}\uparrow} \\\vdots \\\hat{a}_{n_{1}\downarrow} \\\hat{a}_{n_{2}\downarrow} \\\vdots \\\hat{a}^{\dagger}_{n_{1}\uparrow} \\\vdots \\\hat{a}^{\dagger}_{n_{1}\downarrow}\\\vdots\end{array}\right)
\end{equation}
and similarly for $\left(\begin{array}{c}{\boldsymbol{\hat{\gamma}^{\phantom{\dagger}}}} \\\boldsymbol{\hat{\gamma}}^{\dagger}\end{array}\right)$. 
Then the {\it unitary} transformation, Eq.~\eqref{quasi-particle  transformation}, between the two can then be written in block matrix form as
\begin{equation}
\label{matrix form of Bogoliubov transform}
\left(\begin{array}{c}{\boldsymbol{\hat{\gamma}^{\phantom{\dagger}}}} \\\boldsymbol{\hat{\gamma}}^{\dagger}\end{array}\right)=\left(\begin{array}{cc}{\bm U}^{\dagger} & {\bm V}^{\dagger} \\{\bm V}^{\rm T} & {\bm U}^{\rm T}\end{array}\right)\left(\begin{array}{c}{\bm{\hat{a}^{\phantom{\dagger}}}} \\{\bm{\hat{a}}^{\dagger}}\end{array}\right)\equiv\mathbfcal{U}\left(\begin{array}{c}{\bm{\hat{a}^{\phantom{\dagger}}}} \\{\bm{\hat{a}}^{\dagger}}\end{array}\right).
\end{equation}
By unitarity $(\mathbfcal{U}^{\dagger}=\mathbfcal{U}^{-1})$ of the transformation we require
\begin{equation}
\label{Matrix U unitary condition}
\mathbfcal{U}^{\dagger}\mathbfcal{U}=\mathbfcal{U}\mathbfcal{U}^{\dagger}=\openone,
\end{equation}
where
\begin{equation}
\label{calUdagger}
\mathbfcal{U}^{\dagger}=\left(\begin{array}{cc}{\bm U}^{} & {\bm V}^{*} \\{\bm V}^{} & {\bm U}^{*}\end{array}\right). 
\end{equation}
Thus from Eqs.~\eqref{Matrix U unitary condition} and \eqref{calUdagger}  this  implies we must have 
\begin{align}
\label{unitary constraints}
&{\bm U}^{\dagger}{\bm U}+{\bm V}^{\dagger}{\bm V}=\openone,\hspace{1cm}{\bm U}^{}{\bm U}^{\dagger}+{\bm V}^{*}{\bm V}^{\rm T}=\openone\nonumber,\\&
{\bm U}^{\rm T}{\bm V}+{\bm V}^{\rm T}{\bm U}=\mathbb{0},\hspace{1cm}{\bm U}^{}{\bm V}^{\dagger}+{\bm V}^{*}{\bm U}^{\rm T}=\mathbb{0}.
\end{align}
Furthermore, from Eq.~\eqref{matrix form of Bogoliubov transform} the inverse transformation is given as
\begin{equation}
\left(\begin{array}{c}{\bm{\hat{a}^{\phantom{\dagger}}}} \\{\bm{\hat{a}}^{\dagger}}\end{array}\right)=\mathbfcal{U}^{-1}\left(\begin{array}{c}{\boldsymbol{\hat{\gamma}^{\phantom{\dagger}}}} \\\boldsymbol{\hat{\gamma}}^{\dagger}\end{array}\right)=\mathbfcal{U}^{\dagger}\left(\begin{array}{c}{\boldsymbol{\hat{\gamma}^{\phantom{\dagger}}}} \\\boldsymbol{\hat{\gamma}}^{\dagger}\end{array}\right), 
\end{equation}
or explicitly by
\begin{subequations}
\label{mode operators in terms of quasi-particle operators}
\begin{align}
\hat{a}^{}_{\alpha}&=\sum_{\alpha'}\Big(U^{}_{\alpha,\alpha'}\hat{\gamma}^{}_{\alpha'}+V^{*}_{\alpha,\alpha'}\hat{\gamma}^{\dagger}_{\alpha'}\Big),\\
\hat{a}^{\dagger}_{\alpha}&=\sum_{\alpha'}\Big(V^{}_{\alpha,\alpha'}\hat{\gamma}^{}_{\alpha'}
+U^{*}_{\alpha,\alpha'}\hat{\gamma}^{\dagger}_{\alpha'}\Big).
\end{align}
\end{subequations}

Next we  assume a (unnormalized) mean-field ground state of the form $|\Phi\rangle=\prod_{\alpha}\hat{\gamma}^{}_{\alpha}|{\rm vac}\rangle$. The  mean-field ground state acts as a quasi-particle vacuum and satisfies $\hat{\gamma}^{}_{\alpha}|\Phi\rangle=0$ for all $\alpha$.    This implies  $\langle \Phi|\hat{\gamma}^{\dagger}_{\alpha}\hat{\gamma}^{}_{\alpha'}|\Phi\rangle=0$. 
We  further require that there be no residual pairing between quasi-particles in the ground state, i.e., 
\begin{equation}
\langle \Phi|\hat{\gamma}^{}_{\alpha}\hat{\gamma}^{}_{\alpha'}|\Phi\rangle=\langle \Phi|\hat{\gamma}^{\dagger}_{\alpha}\hat{\gamma}^{\dagger}_{\alpha'}|\Phi\rangle=0.
\end{equation}
Thus the one-body density matrix in the single-particle basis is given by
\begin{align}
\label{one-body density matrix}
\rho^{}_{\alpha,\alpha'}&\equiv\frac{\langle \Phi|\hat{a}^{\dagger}_{\alpha}\hat{a}^{}_{\alpha'}|\Phi\rangle}{\langle \Phi| \Phi\rangle}=\sum_{\beta}V^{}_{\alpha,\beta}V^{*}_{\alpha',\beta},\nonumber\\& =\left({\bm V}{\bm V}^{\dagger}\right)_{\alpha,\alpha'}
\end{align}
and similarly   for the anomalous one-body density matrix 
\begin{align}
\label{anomalous pairing matrix}
\kappa^{}_{\alpha,\alpha'}&\equiv\frac{\langle \Phi|\hat{a}^{}_{\alpha}\hat{a}^{}_{\alpha'}|\Phi\rangle}{\langle \Phi|\Phi\rangle}=\sum_{\beta}U^{}_{\alpha,\beta}V^{*}_{\alpha',\beta},\nonumber\\&= \left({\bm U}{\bm V}^{\dagger}\right)_{\alpha,\alpha'}.
\end{align}
One can show that these quantities obey the following symmetry conditions 
\begin{align}
\label{symmetries of rho and kappa}
\boldsymbol{\rho}=\boldsymbol{\rho}^{\dagger}\hspace{1cm}\text{and} \hspace{1cm}\boldsymbol{\kappa}^{*}=-\boldsymbol{\kappa}^{\dagger}. 
\end{align}

Using Eqs.~\eqref{mode operators in terms of quasi-particle operators} the full Hamiltonian, Eq.~\eqref{Full Hamiltonian in composite index notation}, can be expressed in terms of quasi-particle operators as 
\begin{align}
\label{Exact Hamiltonian in the quasi-particle basis}
\hat{H}^{}_{}=E_{0}+\hat{H}^{}_{\rm 1b}+\hat{H}^{}_{\rm 2b},
\end{align}
where
\begin{align}
\label{HFB Ground state energy}
&E^{}_{\rm 0}={\rm Tr}\, \left({\bm T}\boldsymbol{\rho}^{*}-\frac{1}{2}\boldsymbol{\Gamma}\boldsymbol{\rho}^{*}+\frac{1}{2}\boldsymbol{\Delta}\boldsymbol{\kappa}^{*}\right),
\end{align}
is the quasi-particle mean-field ground (vacuum) state energy, and 
 $\hat{H}^{}_{\rm 1b}$ and $\hat{H}^{}_{\rm 2b}$ are one- and two-body quasi-particle operator terms. The one- and two-body terms are not of use in the present work and are quite lengthy, and thus, will not be  explicitly given here.  The inclusion of these terms  would be necessary  for excited quasi-particle states, for a  finite temperature, or the inclusion of quasi-particle interactions, which we do not consider here.  They can be found in the literature, for example, see Ref.~\onlinecite{SignoracciPRC2015}.  
In Eq.~\eqref{HFB Ground state energy} we have also defined the Hartree energy $\Gamma^{}_{\alpha,\delta}$ and pairing matrix $\Delta_{\alpha,\beta}$ as  
 \begin{align}
\Gamma^{}_{\alpha,\delta}&=\sum_{\beta,\gamma}\overline{U}^{}_{\alpha,\beta,\delta,\gamma}\rho^{*}_{\gamma,\beta}\\
\label{pairing matrix}
\Delta_{\alpha,\beta}&=\frac{1}{2}\sum_{\delta,\gamma}\overline{U}^{}_{\alpha,\beta,\delta,\gamma}\kappa^{}_{\delta,\gamma}.
\end{align}

Similarly, the local  spin-resolved densities, magnetization, and  pairing amplitude are given by 
\begin{equation}
\label{local spin density}
n^{}_{\sigma}(x)=\sum_{n,n'}\psi^{*}_{n}(x)\psi^{}_{n'}(x)\rho^{}_{n\sigma,n'\sigma},
\end{equation}
\begin{equation}
\label{local spin mag}
m(x)=n^{}_{\uparrow}(x)-n^{}_{\downarrow}(x),
\end{equation}
and
\begin{equation}
\label{local pairing amplitude}
\Delta^{}_{\sigma,\sigma'}(x)=\sum_{n,n'}\psi^{}_{n}(x)\psi^{}_{n'}(x)\kappa^{}_{n\sigma,n'\sigma'}
\end{equation}
 respectively.

Finally,  we seek to numerically minimize the ground state energy $E_{0}$, Eq.~\eqref{HFB Ground state energy}, as a function of the matrices ${\bm U}$ and ${\bm V}$, subject to the unitary constraints Eqs.~\eqref{unitary constraints}, the symmetry relations of $\boldsymbol{\rho}$ and $\boldsymbol{\kappa}$ given in Eq.~\eqref{symmetries of rho and kappa}, and the total particle number constraint 
\begin{equation}
N={\rm Tr}\, \boldsymbol{\rho}={\rm Tr}\, {\bm V}{\bm V}^{\dagger}.
\end{equation}
In an imbalanced gas with different numbers for  spin-$\uparrow$ and spin-$\downarrow$
\begin{align}
N^{}_{\uparrow}=\sum_{n}\rho_{n\uparrow,n\uparrow}\hspace{1cm}\text{and}\hspace{1cm}N^{}_{\downarrow}=\sum_{n}\rho_{n\downarrow,n\downarrow},
\end{align}
are the relevant constraints. 

Because the ground state energy, Eq.~\eqref{HFB Ground state energy}, is expressed in terms of $\boldsymbol{\rho}$ and $\boldsymbol{\kappa}$ it is  natural to use their matrix elements as the variables of minimization as opposed to the matrix elements of ${\bm U}$ and ${\bm V}$, which appear in the unitary constraints,  Eqs.~\eqref{unitary constraints}.  Using the definitions of $\boldsymbol{\rho}$ and $\boldsymbol{\kappa}$ given in Eq.~\eqref{one-body density matrix} and Eq.~\eqref{anomalous pairing matrix}, one can show that the unitary constraints also be expressed as:
 \begin{subequations}
 \begin{align}
  \boldsymbol{\kappa}^{\rm T}+\boldsymbol{\kappa}&=\mathbb{0},\\
  \boldsymbol{\kappa}\boldsymbol{\rho}-\boldsymbol{\rho}\boldsymbol{\kappa}^{}&=\mathbb{0},\\
  \boldsymbol{\rho}^{2}-\boldsymbol{\rho}-\boldsymbol{\kappa}^{*}\boldsymbol{\kappa}^{}&=\mathbb{0}.
 \end{align}
  \end{subequations}
 Along with the condition $\boldsymbol{\rho}=\boldsymbol{\rho}^{\dagger}$ this completes the minimization constraints.

\section{Results}
\label{Sec: Results}
Our next task is to
apply the HFB formalism outlined in the previous section to scenarios  that are relevant to recent
experiments in  ultracold atomic gases, namely spatially confined  one-dimensional imbalanced spin-\nicefrac{1}{2}   Fermi systems.  In Sec.~\ref{subsection: Application to the Harmonic Trap}, the external trapping potential is chosen to be harmonic and in Sec.~\ref{subsection: Application to the Hard-Wall Trap} a hard-wall box is assumed. For both setups we use a short-ranged particle-particle interaction  $U(x-x')=\lambda \delta(x-x')$, which is relevant  for dilute ultracold atomic gases.   

We will further simplify the problem by assuming both
$\boldsymbol{\rho}$ and $\boldsymbol{\kappa}$ are real. This choice precludes
a complex FF-type pairing amplitude.  Our justification for this assumption is that
the LO-type state (in which pairing is real) is believed to be energetically more stable~\cite{LarkinJETP64}; nonetheless,
relaxing this simplification will be left for future work.  
With this assumption,
the HFB ground state energy and constraint conditions become, 
\begin{align}
\label{real HFB Ground state energy}
&E^{}_{0}={\rm Tr}\, \left({\bm T}\boldsymbol{\rho}^{}-\frac{1}{2}\boldsymbol{\Gamma}\boldsymbol{\rho}^{}+\frac{1}{2}\boldsymbol{\Delta}\boldsymbol{\kappa}^{}\right),
\end{align}
and
\begin{equation}
 N^{}_{\uparrow}=\sum_{n}\rho_{n\uparrow,n\uparrow}\hspace{0.5cm}\text{and}\hspace{0.5cm}N^{}_{\downarrow}=\sum_{n}\rho_{n\downarrow,n\downarrow},
 \end{equation}
 along with
 \begin{subequations}
 \begin{align}  \boldsymbol{\rho}&=\boldsymbol{\rho}^{\rm T},\\
  \boldsymbol{\kappa}^{}&=-\boldsymbol{\kappa}^{\rm T},\\
  \boldsymbol{\kappa}\boldsymbol{\rho}&=\boldsymbol{\rho}\boldsymbol{\kappa}^{},\\
  \boldsymbol{\rho}^{2}-\boldsymbol{\rho}&=\boldsymbol{\kappa}^{}\boldsymbol{\kappa}^{}.
 \end{align}
  \end{subequations}
 
The solution to a large scale constrained non-linear  minimization problem is needed to find the ground state energy.   The  open-source  software IPOPT \cite{IPOPT} was used for this purpose. In principle, both $\boldsymbol{\rho}$ and $\boldsymbol{\kappa}^{}$ are infinite-dimensional matrices, but for computational purposes, a finite representation must be used. In a subspace of the full Hilbert space, each matrix has a linear size of $2{\cal D}$, where the 2 accounts for the spin and ${\cal D}$ is the dimension of the subspace.   For both trapping potentials, we used ${\cal D}=40$.  This amounts to a minimization  problem in approximately 13,000 variables,  the total matrix  elements of $\boldsymbol{\rho}$ and $\boldsymbol{\kappa}$.  With this cutoff, the number of particles, and interaction strength $\lambda$ used in the calculations, the occupation of the highest state in the restricted Hilbert space remains only at the 1\% level, for either trapping potential.

\begin{figure*}
\subfloat[$N_{\uparrow}=N_{\downarrow}=20$ $(P=0)$\label{fig:fig1a}]{%
  \includegraphics[width=0.49\textwidth]{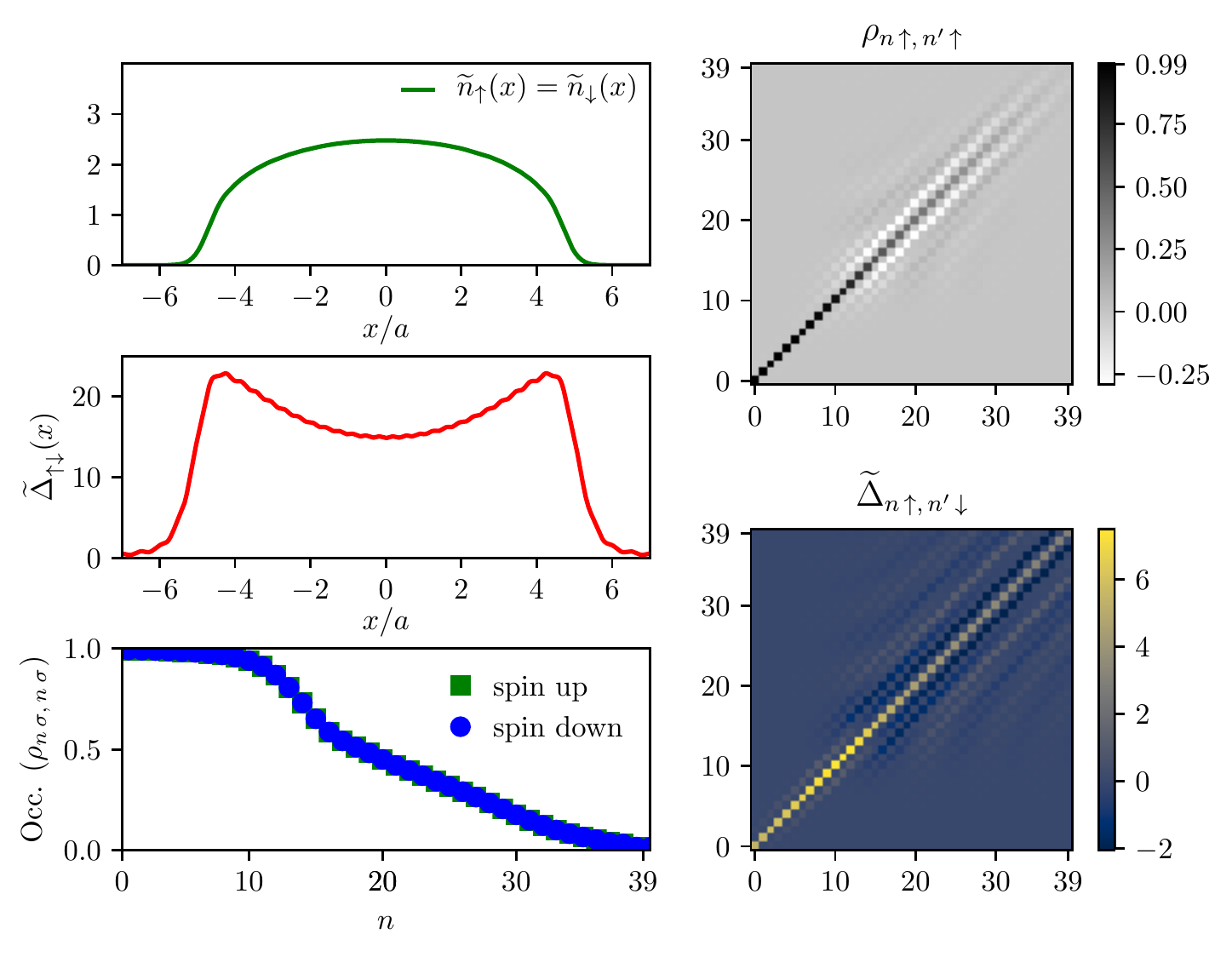}%
}\hfill
\subfloat[$N_{\uparrow}=20$ and $ N_{\downarrow}=18$ $(P\approx 0.05)$\label{fig:fig1b}]{%
  \includegraphics[width=.49\textwidth]{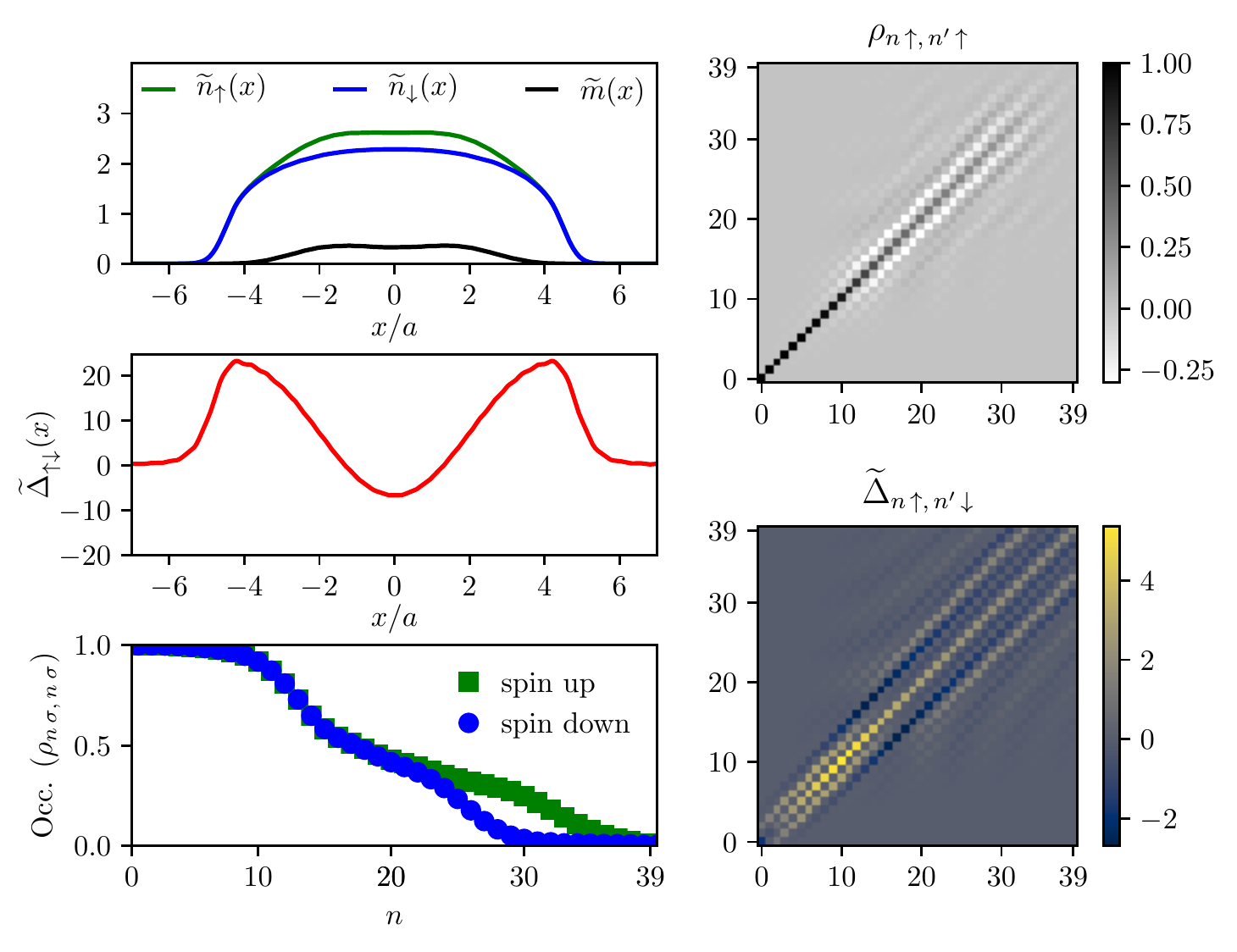}%
}\hfill
\vspace{-0.25cm}
\subfloat[$N_{\uparrow}=20$ and $N_{\downarrow}=17$ $(P\approx 0.08)$\label{fig:fig1c}]{%
  \includegraphics[width=.49\textwidth]{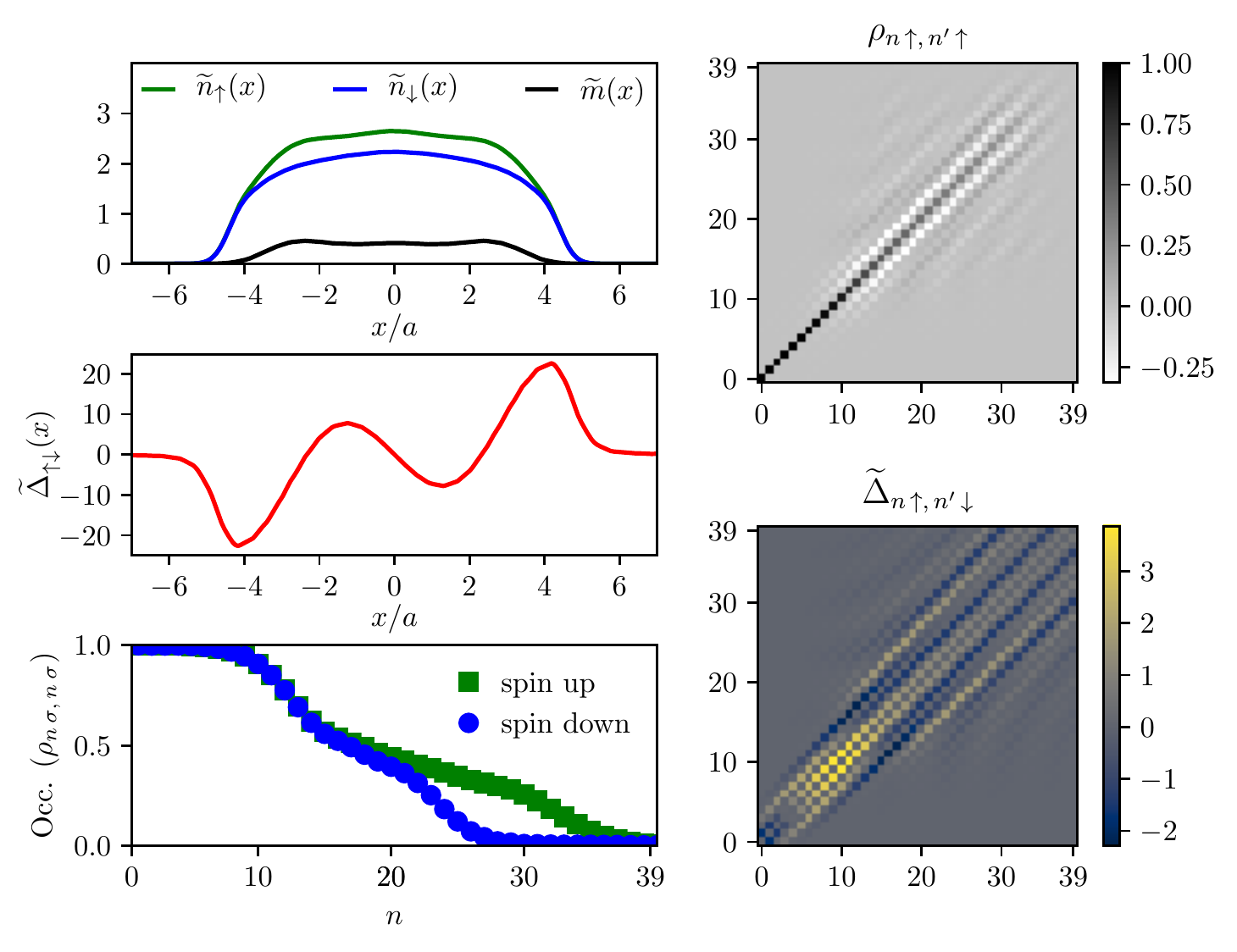}%
}\hfill
\subfloat[$N_{\uparrow}=20$ and $ N_{\downarrow}=16$ $(P\approx 0.11)$\label{fig:fig1d}]{%
  \includegraphics[width=.49\textwidth]{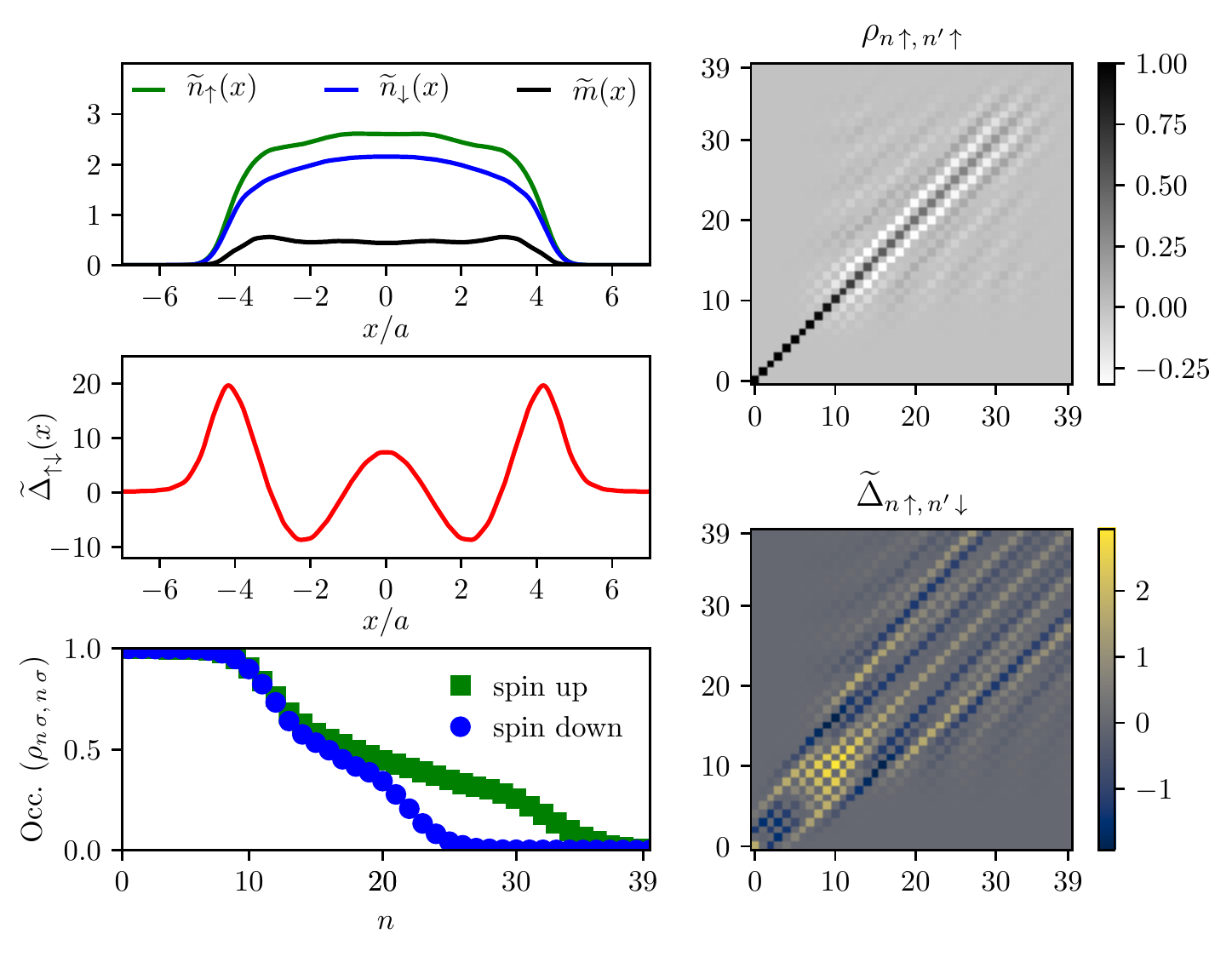}%
}
\caption{(Color online) Harmonic trap case: In each panel,
  Figs.~\ref{fig:fig1a}-\ref{fig:fig1d}, the dimensionless form of
  the ground state  local spin-resolved densities
  $\tilde{n}^{}_{\sigma}(x)=a\, n_{\sigma}(a\,\tilde{x})$,
  magnetization
  $\tilde{m}(x)=\tilde{n}^{}_{\uparrow}(x)-\tilde{n}^{}_{\downarrow}(x)$,
  and pairing amplitude
  $\tilde{\Delta}^{}_{\sigma,\sigma'}(x)=a\,\omega^{-1}_{0}\,\Delta^{}_{\sigma,\sigma'}(a\,\tilde{x})$
  for a harmonic trapping potential are shown for various spin
  polarizations
  $P=(N_{\uparrow}-N_{\downarrow})/(N_{\uparrow}+N_{\downarrow})$,
  where $\omega_{0}$ and $a$ are the harmonic trap frequency and
  oscillator length respectively.  In addition, for each polarization
  the one-body density matrix $\rho_{\alpha,\beta}$, the occupations
  of the harmonic states $\rho_{\alpha,\alpha}$, and pairing matrix
  $\tilde{\Delta}_{\alpha,\beta}=\omega^{-1}_{0}\,\Delta_{\alpha,\beta}$
  are also shown. The dimensionless interaction strength for all plots
  is $\tilde{\lambda}=(\omega_{0}a)^{-1}\lambda= -100/\pi^{2}$.}
\label{fig:fig1}
\end{figure*}

\subsection{Harmonic Trap}
\label{subsection: Application to the Harmonic Trap}
In this section we specialize the general formalism of Sec.~\ref{Sec: Theory} to the case of a harmonically trapped gas with a short-ranged two-body interaction.  The external trapping potential  of a 1D harmonic oscillator with trapping frequency $\omega_{0}$ is taken to be
\begin{equation}
\label{Harmonic trapping potential}
V(x)=\frac{1}{2}m\,\omega^{2}_{0}x^{2},
\end{equation}
and  single-particle wave functions of $V(x)$ are the well-known harmonic oscillator states 
\begin{equation}
\psi^{}_{n}(x)=\frac{1}{\sqrt{2^{n}n!a\sqrt{\pi}}}e^{-x^{2}/(2a^{2})}H^{}_{n}(x/a),
\end{equation}
where $H^{}_{n}(x)$ are the Hermite  polynomials and $a=(m\omega_{0})^{-1/2}$ is the oscillator length. The allowed principal quantum numbers are  $n=0,1,2,3,\ldots$, and the single-particle spectrum is  $\epsilon_{n}=\omega_{0}(n+1/2)$. 
Using the above  oscillator states and  a delta function interaction, the matrix elements of the two-body interaction term, Eq.~\eqref{pseudo two-body interaction matrix element}, are given by 
\begin{widetext}
 \begin{align}
\label{Harmonic interaction matrix elements}
U^{}_{n_{1},n_{2},n_{3},n_{4}}&=\lambda\int\limits_{-\infty}^{\infty}{\rm d}x\,\psi^{*}_{n_{1}}(x)\psi^{*}_{n_{2}}(x)\psi^{}_{n_{3}}(x)\psi^{}_{n_{4}}(x)\nonumber\\&=
 \begin{cases}
    0  \hspace{0.5cm}\text{if $2M$ is odd} \\
   \displaystyle{\frac{\lambda}{\pi a}(-1)^{M-n_{3}-n_{1}}2^{-1/2}(n_{1}!n_{2}!n_{3}!n_{4}!)^{-1/2}\frac{\Gamma(M-n_{2}+1/2)\Gamma(M-n_{4}+1/2)}{\Gamma(M-n_{2}-n_{4}+1/2)}}\\ \hspace{1cm}\times {}_{3}F_{2}(-n_{1},-n_{3},-M+n_{2}+n_{4}+1/2;-M+n_{4}+1/2,-M+n_{2}+1/2;1)\hspace{0.5cm}  \text{if $2M$ is even},  \\
  \end{cases}
\end{align}
\end{widetext}
where $2M=n_{1}+n_{2}+n_{3}+n_{4}$, $\Gamma(x)$ is the standard gamma function, and ${}_{3}F_{2}(a_{1},a_{2},a_{3};b_{1},b_{2};z)$ is a hypergeometric function. In the evaluation of Eq.~\eqref{Harmonic interaction matrix elements} one needs to integrate
the product of four Hermite  polynomials times a Gaussian. This integral can be found in Ref.~\onlinecite{RDLord1949}.

The four panels of Fig.~\ref{fig:fig1} depict our results for
the evolution of a harmonically trapped 1D Fermi gas for four different values of
the total polarization $P=(N_{\uparrow}-N_{\downarrow})/(N_{\uparrow}+N_{\downarrow})$ at fixed
interaction strength.  Each panel shows
the local spin-resolved densities $n_{\sigma}(x)$ (Eq.~\eqref{local spin density}),
the local magnetization $m(x)$ (Eq.~\eqref{local spin mag}),
the local pairing amplitude $\Delta^{}_{\sigma,\sigma'}(x)$
the one-body mode-resolved density matrix $\rho_{\alpha,\beta}$ (Eq.~\eqref{one-body density matrix}),
and the pairing matrix $\Delta_{\alpha,\beta}$ (Eq.~\eqref{pairing matrix}).

As we have discussed, our primary interest is  how FFLO pairing correlations emerge with increasing $P$ and
how they would be reflected in the local densities.   We start with the balanced case, $P=0$, shown
in Fig.~\ref{fig:fig1a}.  We find a local density and pairing amplitude that are  
spatially inhomogeneous due to the imposed trapping potential, with shapes that are consistent with
earlier work based on mean-field theory and Bethe ansatz~\cite{Kudla15}. Furthermore,
the one-body density and pairing matrices are mostly diagonal
in index space.

Figures \ref{fig:fig1b}-\ref{fig:fig1d} show the evolution of these system properties with increasing $P$,
as homogenous pairing is interrupted by the imposed population imbalance.  Consistent with early
experiment and theory work \cite{LiaoNature2010}, we find the imposed population imbalance leads to a
magnetized central region reflecting a magnetized core and balanced superfluid edges.  With increasing
$P$, the magnetized core increases in size until a critical polarization $P_{\rm c}$ is reached,
beyond which the  entire cloud is polarized. For the
calculations
presented here, we find that $P_{\rm c}$ lies  between 0.11 and 0.14.  This is in close agreement with experiments that find a $P_{\rm c}\sim 0.13$ and in  qualitative agreement with BA+LDA, which predicts a $P_{\rm c}\sim  0.17$ \cite{LiaoNature2010}.

Figures \ref{fig:fig1b}-\ref{fig:fig1d}
also show that the local pairing amplitude $\Delta_{\uparrow\downarrow}(x)$ is
oscillatory in real space in the imbalanced regime, qualitatively consistent with a LO-like
pairing function.   A key well-known
property of the FFLO state is the prediction that the FFLO wave vector
$Q$ is proportional to the imbalance.  Here, this is reflected in the
fact that the number of nodes in $\Delta_{\uparrow\downarrow}(x)$
increases with increasing $P$. In fact, for the results presented here the number of nodes
is precisely $N_\uparrow - N_\downarrow$, since $\Delta_{\uparrow\downarrow}(x)$ shows
2, 3, and 4 nodes in Figs.~\ref{fig:fig1b}-\ref{fig:fig1d}, respectively. 
Thus,  for each additional
``unpaired'' majority spin another node in the pairing function is
created.

Thus, our results confirm the expectation of an FFLO phase of harmonically-trapped 1D imbalanced
Fermi gases, with nodes in the local pairing amplitude.  Our next question is how these
nodes are reflected in the principal
observable in cold atom experiments, i.e., the local atom densities.  Unfortunately,
as seen in Figs.~\ref{fig:fig1b}-\ref{fig:fig1d}, the location of the nodes
is only weakly reflected in the local densities and magnetization.
Even if these small oscillations in the magnetization could be measured,
providing a possible signature of the FFLO phase in experiments,
we must recall that our results are for the ground state.  
Finite temperature effects would most certainly suppress these
small oscillations even further, making them likely
undetectable~\cite{LiuPRA2008} via a direct imaging of the
density.  Extending these results  to finite temperature
\cite{ALGoodmanNuclerPhysA1981} will be left for future work.

The nontrivial off-diagonal values seen in the one-body density matrix
$\rho_{\alpha,\beta}$ and the pairing matrix
$\Delta_{\alpha,\beta}$ signify 
scattering  from the Hartree potential, and more importantly they indicate the fact that  the
canonically paired states are not  simple pairs of harmonic oscillator states,
i.e., $\Delta_{n,n'}\not\propto \langle \hat{a}^{}_{n\uparrow}
\hat{a}^{}_{n\downarrow}\rangle\delta^{}_{n,n'}$.  But by the Bloch-Messiah-Zumino theorem \cite{BlochMessiah, Zumino} a  transformation exists that can bring both $\rho_{\alpha,\beta}$ and $\kappa_{\alpha,\beta}$ into diagonal and canonical form respectively.   Thus the Copper paired states can be represented as a linear combination of many harmonic states. Furthermore, with
increasing polarization, the off-diagonal  spectral weight increases,
with no clearly discernible pattern.  The nontrivial occupation
probabilities  $\rho_{\alpha,\alpha}$   of the oscillator states is
also indicative of an unusual pairing solution. 

\begin{figure*}
\subfloat[$N_{\uparrow}=N_{\downarrow}=20$ $(P=0)$\label{fig:fig2a}]{%
  \includegraphics[width=0.49\textwidth]{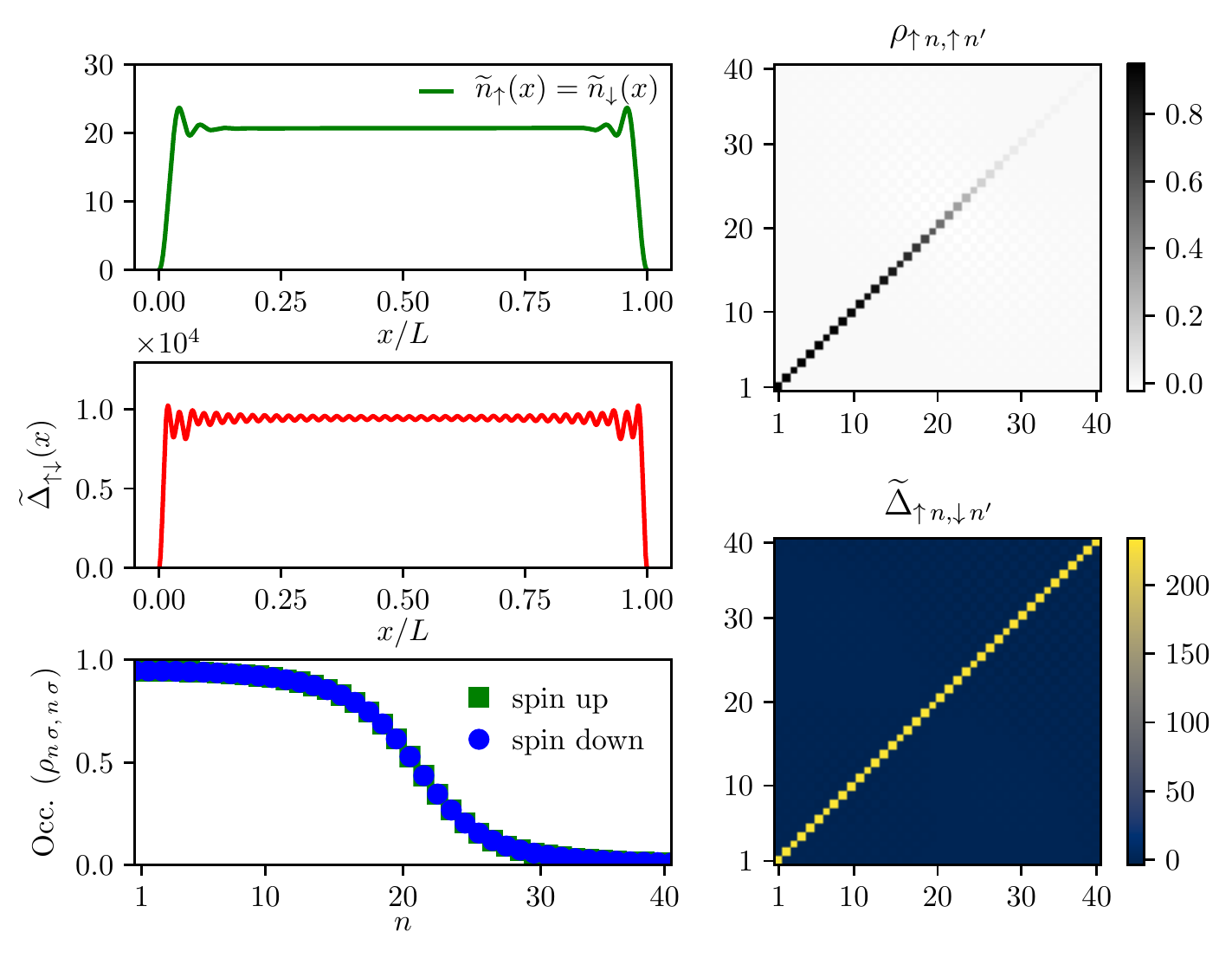}%
}\hfill
\subfloat[$N_{\uparrow}=20$ and $ N_{\downarrow}=19$ $(P\approx 0.025)$\label{fig:fig2b}]{%
  \includegraphics[width=.49\textwidth]{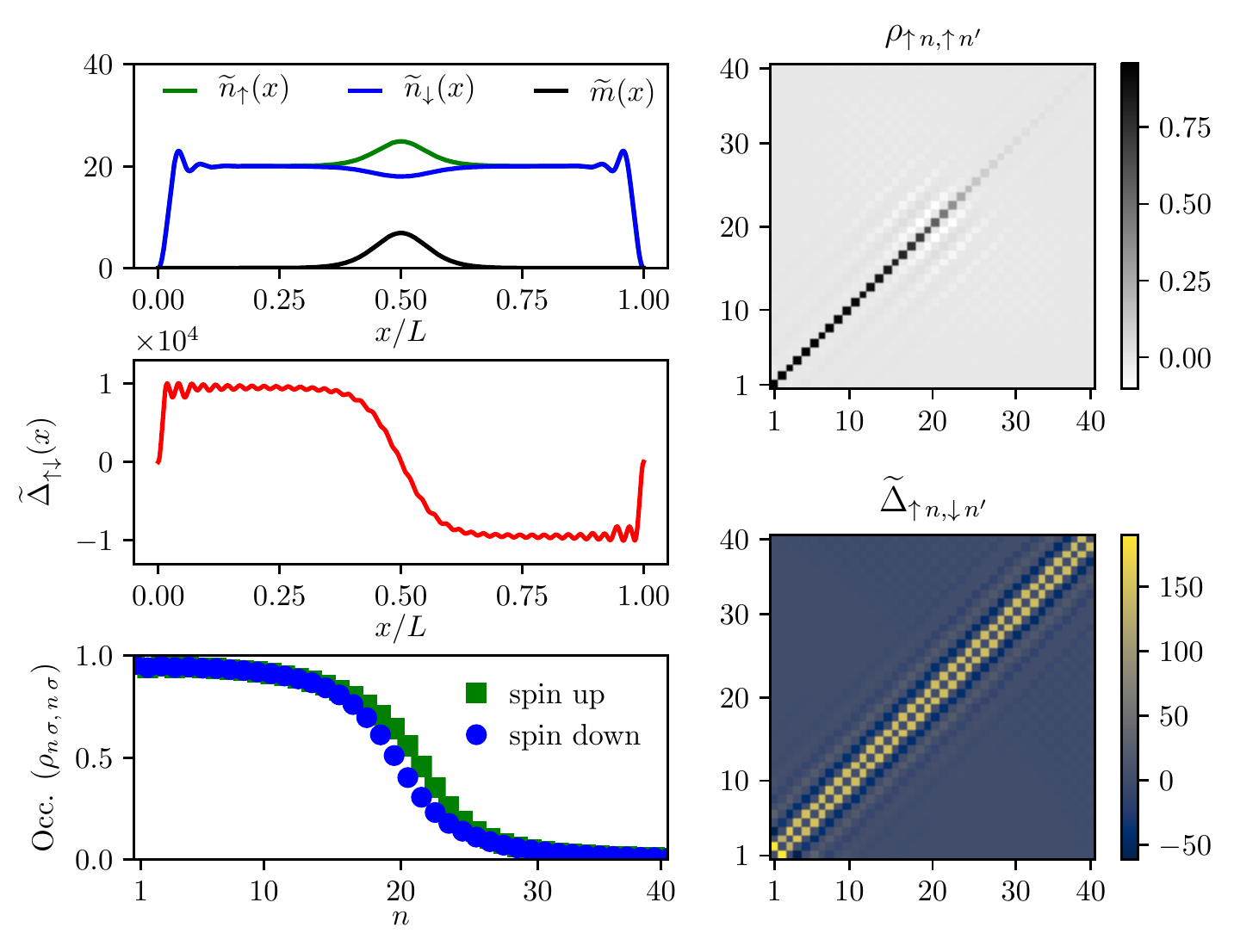}%
}\hfill
\vspace{-0.25cm}
\subfloat[$N_{\uparrow}=20$ and $N_{\downarrow}=18$ $(P\approx 0.05)$\label{fig:fig2c}]{%
  \includegraphics[width=.49\textwidth]{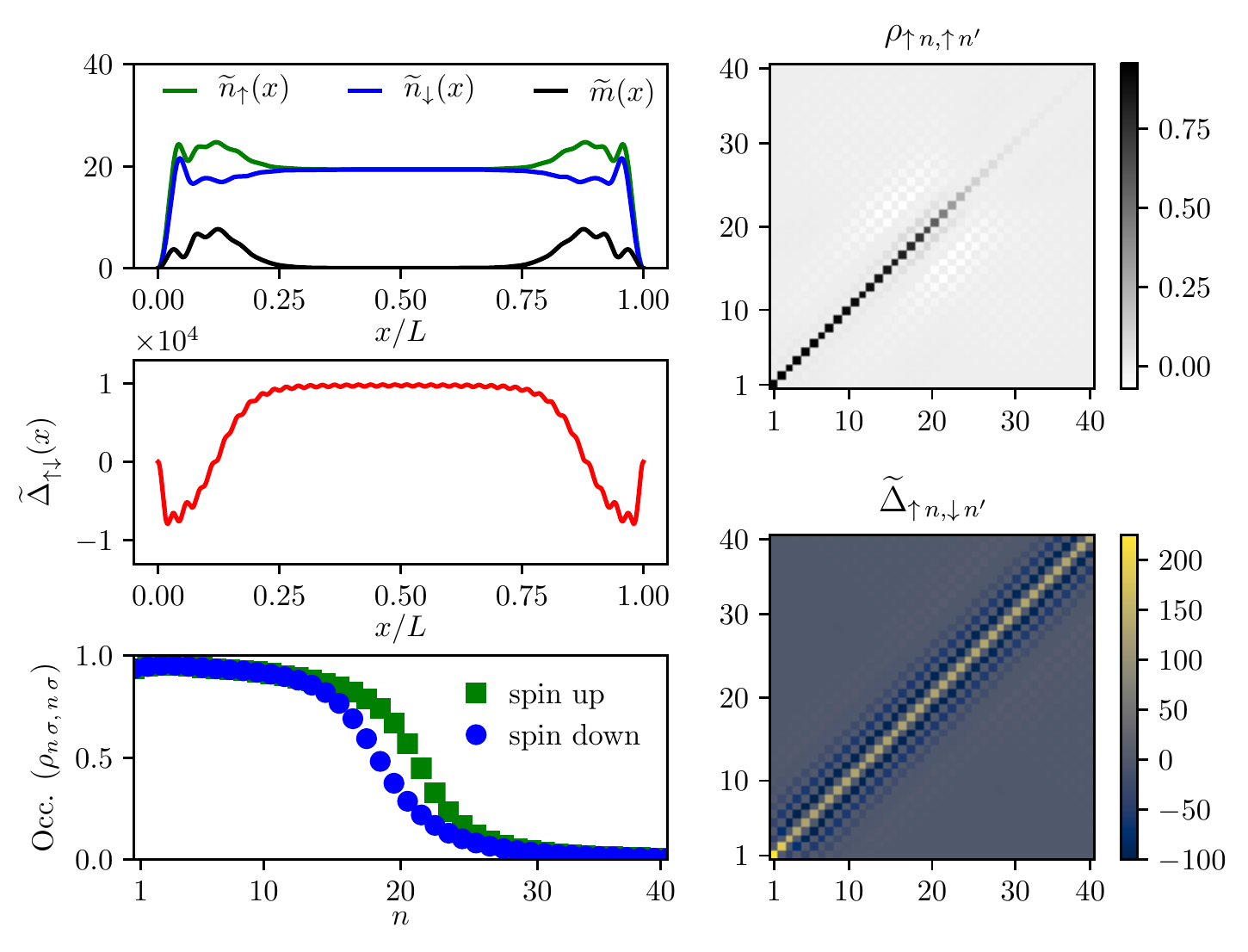}%
}\hfill
\subfloat[$N_{\uparrow}=20$ and $ N_{\downarrow}=17$ $(P\approx 0.08)$\label{fig:fig2d}]{%
  \includegraphics[width=.49\textwidth]{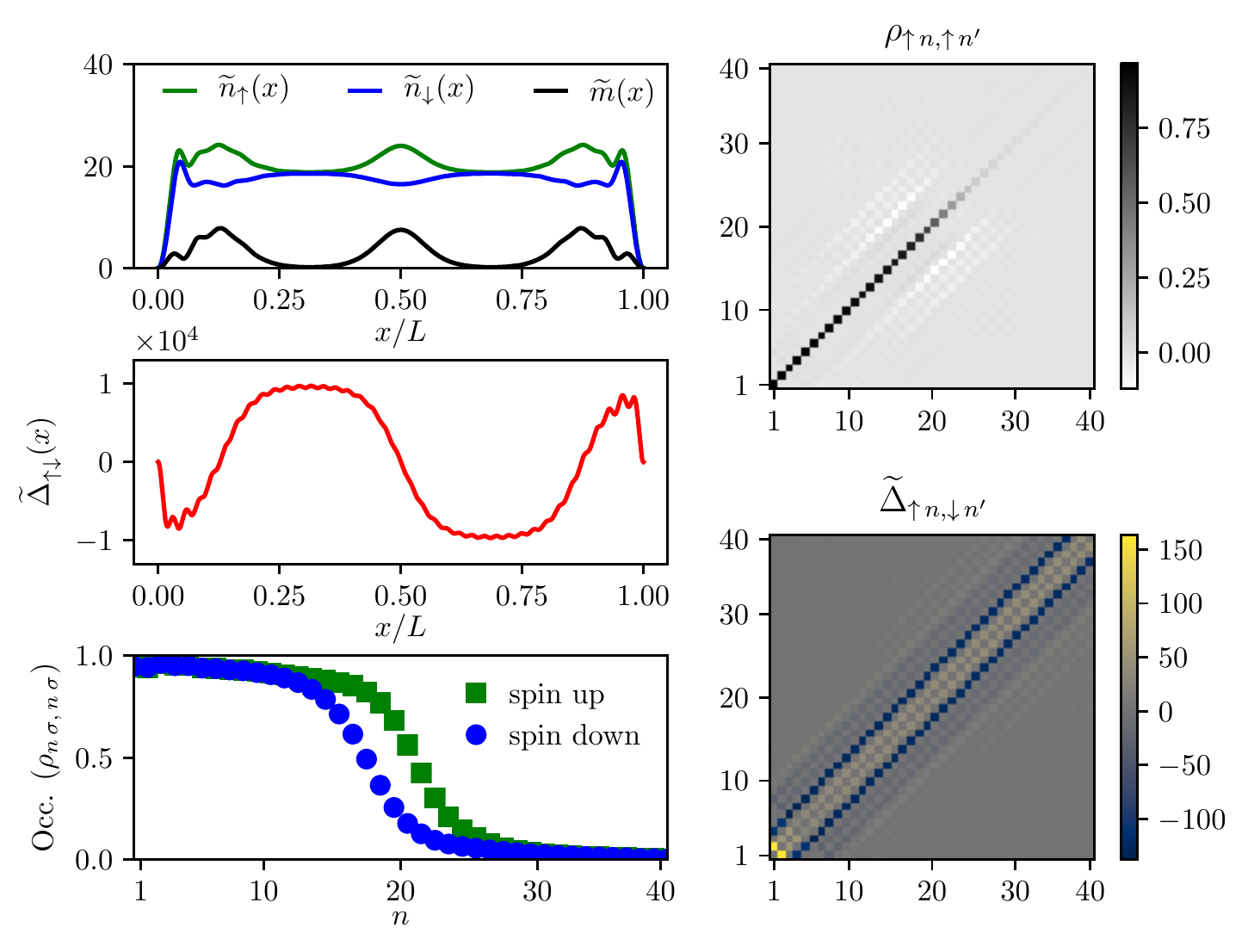}%
}
\caption{(Color online) Hard-wall case: In each panel (a-d), the dimensionless form of  the ground state  local spin-resolved densities $\tilde{n}^{}_{\sigma}(x)=L\, n_{\sigma}(L\,\tilde{x})$, magnetization $\tilde{m}(x)=\tilde{n}^{}_{\uparrow}(x)-\tilde{n}^{}_{\downarrow}(x)$, and pairing amplitude $\tilde{\Delta}^{}_{\sigma,\sigma'}(x)=L\,\varepsilon^{-1}\,\Delta^{}_{\sigma,\sigma'}(L\,\tilde{x})$ for a hard-wall trapping potential are shown for various spin imbalances $P=(N_{\uparrow}-N_{\downarrow})/(N_{\uparrow}+N_{\downarrow})$, where $\varepsilon=\pi^{2}/(2mL^{2})$, and $L$ is the box length.  In addition for each polarization the one-body density matrix $\rho_{\alpha,\beta}$ and pairing matrix $\tilde{\Delta}_{\alpha,\beta}=\varepsilon^{-1}\,\Delta_{\alpha,\beta}$ are also shown. The dimensionless interaction strength for all plots is $\tilde{\lambda}=(\varepsilon L)^{-1}\lambda= -200/\pi^{2}$.}
\label{fig:fig2}
\end{figure*}

\subsection{Hard-Wall Trap}
In this section we turn our attention to the case of a hard-wall or ``box'' shaped
trap in one spatial dimension, characterized by the following
single-particle potential for a 
1D infinite square well of width $L$:
\begin{equation}
\label{hard wall potential}
  V(x) = 
  \begin{cases}
    0 & \text{for } 0\leq x \leq L \\
   \infty & \text{otherwise.}  \\
  \end{cases}
\end{equation}  
The single-particle spectrum of the well is 
\begin{align}
\epsilon_{n}=\frac{\pi^{2}}{2mL^{2}}n^{2}\equiv\varepsilon n^{2},
\end{align}
where $\varepsilon =\pi^{2}/(2mL^{2}) $, which has units of energy
($\hbar =1$).  Numerical results for this system will be shown using
$\varepsilon$ for an energy scale  and $L$ for length.  The
single-particle states for the  potential given by  Eq.~\eqref{hard wall
  potential} are the conventional
\begin{equation}
\psi^{}_{n}(x)=
 \begin{cases}
    \sqrt{\frac{2}{L}}\sin(k_{n}x) & \text{for } 0\leq x \leq L \\
   0 & \text{otherwise,}  \\
  \end{cases}
\end{equation}
where $k_{n}=\pi n/L$ with $n=1,2,3,4,\ldots$.
For a short-range interaction, $U(x-x')=\lambda \delta(x-x')$,
the matrix elements of the two-body interaction term, Eq.~\eqref{pseudo two-body interaction matrix element}, are then
\begin{align}
&U^{}_{n_{1},n_{2},n_{3},n_{4}}=\lambda\int\limits_{0}^{L}{\rm d}x\,\psi^{*}_{n_{1}}(x)\psi^{*}_{n_{2}}(x)\psi^{}_{n_{3}}(x)\psi^{}_{n_{4}}(x),\nonumber\\ &= \frac{\lambda}{4L}\sum_{\substack{\varepsilon_{1},\varepsilon_{2}=\pm\\ \varepsilon_{3},\varepsilon_{4}=\pm}}\varepsilon_{1}\varepsilon_{2}\varepsilon_{3}\varepsilon_{4}\, I(\varepsilon_{1}n_{1}+\varepsilon_{2}n_{2}+\varepsilon_{3}n_{3}+\varepsilon_{4}n_{4}),
\end{align}
where $I(n)$ acts like a Kronecker delta function: 
\begin{equation}
I(n)=
\begin{cases}
1, \text{ if } n=0\\
0, \text{ if } n \neq 0.
\end{cases}
\end{equation}
\label{subsection: Application to the Hard-Wall Trap}

Our results for the hard-wall case are shown in Fig.~\ref{fig:fig2},
where we show the same system properties as in Fig.~\ref{fig:fig1}  for four values of the
imposed polarization $P$.  Starting with panel (a), $P= 0$, we find
 essentially homogeneous results for the local density
and local pairing amplitude, except for small oscillations near the box
edge.  Our results for the
mode resolved one-body density matrix and pairing matrix of this balanced
1D gas are approximately diagonal, with no unusual structure.

Figures \ref{fig:fig2b}-\ref{fig:fig2d} show the fate of this balanced superfluid
state under an imposed population imbalance.  As in the case of a harmonically
trapped gas, such an imbalance leads to a spatially oscillating pair amplitude 
$\Delta^{}_{\sigma,\sigma'}(x)$, with the number of nodes in the cloud interior
being again proportional to $N_\uparrow - N_\downarrow$ and given by $1,2,3$ in
Figs.~\ref{fig:fig2b}, \ref{fig:fig2c}, and \ref{fig:fig2d} respectively. (In this counting we are ignoring the  nodes
of $\Delta^{}_{\sigma,\sigma'}(x)$ at the edges of the box.)

Although the local pairing in the FFLO state is qualitatively similar
to the harmonic case, the reflection of this state in the local atom
densities is strikingly different.  Indeed, Figs.~\ref{fig:fig2b}-\ref{fig:fig2d}
show relatively narrow peaks in the local density of spin-$\uparrow$
atoms (and corresponding small dips in the local spin-$\downarrow$ density)
located spatially near the nodes in $\Delta^{}_{\sigma,\sigma'}(x)$.

Another key difference relative to the harmonic case is that
the local magnetization does not exhibit a central polarized region.
Instead, the magnetized regions are  localized near the
nodes of pairing amplitude, with near-zero net magnetization between
the nodes. Indeed, this FFLO state can be interpreted
in terms of well-formed domain walls in the pairing amplitude,
with excess spins-$\uparrow$ congregating near the nodes,
providing a robust signature of the FFLO phase of a 1D imbalanced gas in a box trap.

Like the harmonic potential, the  mode-resolved one-body density
$\rho_{\alpha,\beta}$  and  pairing  $\Delta_{\alpha,\beta}$  matrices
process small but nontrivial off-diagonal spectral weight. Again, this
implies that the  Cooper pairs of the system  are  a  combination of
many single-particle states.  

Although the off-diagonal terms of $\rho_{\alpha,\beta}$ are small, these terms are ultimately responsible for the FFLO modulation seen in the density. For example, if the density matrix was diagonal, $\rho_{\alpha,\beta}\propto \delta_{\alpha,\beta}$, then the local density (Eq.~\eqref{local spin density}), would reduce to $n_{\sigma}(x)= \sum_{n} |\psi^{}_{n}(x)|^{2}\rho_{n\sigma,n\sigma}$. Then from  Fig.~\ref{fig:fig2} one can see that  the occupations probabilities $\rho_{n\sigma,n\sigma}$ are qualitatively similar to what one would expect to see for a non-interacting Fermi gas at some non-zero temperature;  thus the density would simply resemble a non-interacting system at a finite temperature and  be devoid of any FFLO oscillations.     

In this geometry, the  off-diagonal terms of the pairing matrix also have a much clearer physical interpretation: the states $n$ and $n'$ that are generally the most strongly paired are those that differ by the particle number imbalance. For example, when $N_{\uparrow}-N_{\downarrow}=3$, shown in Fig.~\ref{fig:fig2d}, the states labeled by $n=n'\pm3$   show the strongest correlations.  This implies  that under an imbalance, a majority spin in state $n$  tends to most strongly pair with a minority spin in state $n'=n-(N_{\uparrow}-N_{\downarrow})$.   This is in agreement with the ansatz put-forth by the present authors  in Ref.~\cite{PattonPRA2017} for a harmonic trap. Here, one imagines that  a majority spin at its Femi surface $\epsilon^{\rm F}_{N_{\uparrow}}$ will  form a pair  with a minority spin on the other Fermi surface $\epsilon^{\rm F}_{N_{\downarrow}}$.  

\section{Discussion and Conclusions}
\label{Sec: Discussion and Conclusions}

The HFB ground state $|{\Phi}\rangle$, which is also the vacuum state of the generalized Bogoliubov quasi-particles $|{\Phi}\rangle =\prod_{\alpha}\hat{\gamma}^{}_{\alpha}|{\rm vac}\rangle$,
can also be, by Thouless's theorem \cite{SchuckBook},   expressed as
\begin{equation}
\label{Thouless theorem}
|{\Phi}\rangle\propto \exp\left(\frac{1}{2}\sum_{\alpha,\beta}Z^{}_{\alpha,\beta}\hat{a}^{\dagger}_{\alpha}\hat{a}^{\dagger}_{\beta}\right)|{\rm vac}\rangle,
\end{equation}
where  ${\bm Z}=({\bm V}{\bm U}^{-1})^{\dagger}=(\boldsymbol{\rho}\boldsymbol{\kappa}^{-1})^{\dagger}$ is an anti-symmetric  matrix that gives the probability amplitude of creating a Copper pair comprised  of states $\alpha$ and $\beta$. Thus in generality state $\alpha$ could form Cooper  pairs with many  other states, instead of only a single one, as is assumed in  the standard BCS wave function, and therefore in the presence of a population imbalance all atoms can still take part in the superfluid state.  For the trapped and imbalanced systems considered here,  we have found that it is vitally important to account for all possible pairing Hartree-Fock correlations, especially in the harmonic system.

Our analysis shows that while both harmonic and box-shaped traps can host FFLO-like states
in the 1D regime, the signature of this state in the harmonic case in the local density
is extremely weak and probably unobservable in a real experiment at nonzero temperature.
This suggests a competition between the spatial variation of the FFLO state and
the slowly-varying potential of the harmonic trap.  
For the case of a 1D box-shaped trap, however, our results show a striking signature of
the FFLO phase in the density profile, indicating that spin-sensitive measurements will
be able to discern this exotic state.

Future work will involve generalizing these results to non-zero
temperature, which would involve solving a generalized gap equation
for the quasi-particle energy spectrum.  This will be essential to determine
whether the sharp signatures of the FFLO state in the box trap survive
nonzero temperature.
Additional future work will extend this
formalism to higher dimensions to more generally understand how
the spatial profile of the confining potential affects the FFLO phase
of an imbalanced Fermi gas and its observability in experimental observables
such as the local density.

\acknowledgements
KRP would like the thank Georgia Southern University for generous startup funding, which contributed to this work.


\begin{thebibliography}{99}
\bibitem{footnote} Note that most references dealing with HFB theory do not express the Hamiltonian using Eq.~\eqref{pseudo two-body interaction matrix element}, but instead use the two-body matrix element $U^{\rm 2b}_{\alpha,\beta,\delta,\gamma}=\left<\alpha,\beta | U |\delta,\gamma \right> = \int{\rm d}x{\rm d}x'\, \psi^{*}_{\alpha}(x)\psi^{*}_{\beta}(x')U(x-x')\psi^{}_{\delta}(x)\psi^{}_{\gamma}(x')=U^{}_{\alpha,\beta,\gamma,\delta}$, which differs from Eq.~\eqref{pseudo two-body interaction matrix element} by the ordering of the last two indices. This difference can lead to minus signs or operator ordering differences from expressions found in the literature when compared to Eq.~\eqref{Exact Hamiltonian: composite index}. Nonetheless we  stick with using Eq.~\eqref{pseudo two-body interaction matrix element} as it is the most common expression in the ultracold atomic gas community.
\end{thebibliography}
\end{document}